\documentclass[twocolumn,tighten,times]{aastex631}

\newcommand{\um}{$\mu$m}
\definecolor{navyblue}{RGB}{0,50,250}

\definecolor{firebrick}{RGB}{208,34,34}

\definecolor{green}{RGB}{34,208,70}

\shorttitle{Probing the dust grain alignment mechanisms in galaxies}
\shortauthors{Lopez-Rodriguez, E.}
\graphicspath{{./}{figures/}}

\begin{document}

\title{Probing the dust grain alignment mechanisms in spiral galaxies with M\,51 as the case study}

\author[0000-0001-5357-6538]{Enrique Lopez-Rodriguez}
\affiliation{Kavli Institute for Particle Astrophysics \& Cosmology (KIPAC), Stanford University, Stanford, CA 94305, USA}

\author[0000-0002-6488-8227]{Le Ngoc Tram}
\affil{Max-Planck-Institut f\"ur Radioastronomie, Auf dem H\"ugel 69, 53121 Bonn, German}



\begin{abstract}
Magnetic fields (B-fields) in galaxies have recently been traced using far-infrared and sub-mm polarimetric observations with SOFIA, JCMT, and ALMA. The main assumption is that dust grains are magnetically aligned with the local B-field in the interstellar medium (ISM). However, the range of physical conditions of the ISM, dust grain sizes, and B-field strengths in galaxies where this assumption is valid has not been characterized yet. Here, we use the well-studied spiral galaxy M\,51 as a case study. We find that the timescale for the alignment mechanism arising from magnetically aligned dust grains (B-RAT) dominates over other alignment mechanisms, including radiative precession (k-RAT), and mechanical alignment (v-MAT), as well as to the randomization effect (gas damping). 
We estimate the sizes of the aligned dust grain to be in the range of $0.009-0.182$ \um~and $0.019-0.452$ \um~for arms and inter-arms, respectively. We show that the difference in the polarization fraction between arms and interarms can arise from the enhancement of small dust grain sizes in the arms as an effect of the grain alignment disruption (RAT-D).  We argue that the RAT-D mechanism needs to have additional effects, e.g., intrinsic variations of the B-field structure and turbulence, in the galaxy's components to fully explain the polarization fraction variations within the arms and inter-arms. 

\end{abstract}





\section{Introduction} \label{sec:intro}

Far-infrared (FIR; $50-220$ \um) and sub-mm ($850-1200$ \um) imaging polarimetric observations have opened a new window to explore extragalactic magnetism \citep[e.g.,][]{ELR2018,Jones2019,ELR2020,Jones2020,ELR2021a,ELR2021b,Borlaff2021,Pattle2021b,SALSAIV,SALSAV,ELR2023}. All observed nearby ($\leq20$ Mpc) galaxies (e.g., spirals, mergers, starbursts, active galactic nuclei) show kpc-scale ordered magnetic fields (B-fields) cospatial with the dense and cold phase of the interstellar medium (ISM) and star-forming regions. These observations were performed using the High-Angular Wideband Camera Plus (HAWC+) on board the 2.7-m Stratospheric Observatory for Infrared Astronomy (SOFIA), POL-2 on the James Clerk Maxwell Telescope (JCMT), and the Atacama Large Millimeter/submillimeter Array (ALMA). The angular resolutions of single-dish telescopes are in the range of $5-18$\arcsec, which corresponds to spatial scales of $\sim300$ pc for a typical nearby galaxy at $\sim10$ Mpc and at an angular resolution of $10$\arcsec\ \citep{SALSAIV}, while ALMA has performed polarimetric observations of resolution of $\sim5$ pc in the starburst galaxy NGC\,253 at a distance of $3.50$ Mpc \citep{ELR2023}. Furthermore, recent ALMA polarimetric observations have measured B-fields of $\sim5$ kpc scale in a gravitationally lensed dusty star-forming galaxy at redshift $z=2.6$ by means of polarized thermal emission in the rest frame of $\sim350$ \um\ \citep{Geach2023}. These measurements rely on that the dust grains are magnetically aligned with the local B-field \citep[e.g.,][]{Andersson2015}, and that the measured B-field orientation is a density-weighted average of a non-trivial B-field structure within the several hundred pc size given by the angular resolution \citep{MA2023}. Thus, several fundamental questions arise: are the FIR/sub-mm polarimetric observations measuring magnetically aligned dust grains, or are these observations affected by other dust alignment mechanisms? Is the dependence of the polarization with total intensity due to a loss of dust grain alignment efficiency or intrinsic variations of the B-field?

Theories of grain alignment mechanisms have been in active development since the first discovery of the polarization signature of aligned non-spherical dust grains in the interstellar medium (ISM) by \citet{Hiltner1949} and \citet{Hall1949}. The paramagnetic relaxation proposed by \citet[DG;][]{DG1951} had been considered as a classical mechanism. However, paramagnetic relaxation has been shown to be inefficient as the rotation of large grains can be easily randomized and then weakly aligned \citep{HL2016a,HL2016b}, which is opposite to observations. \citet{Purcell1979} showed that the grains could suprathermally rotate with systematic torques, which is due to the desorption of the recoil of H$_{2}$ molecules from the grain surface to the gas phase. This suprathermal rotation could avoid being randomized by the gas random collision. However, it is fixed to the grain body, which is essentially affected by even a small change in the temperature and radiation in the medium. \citet{LD1999} demonstrated that rapid thermal flipping leads to a systematic torque to a zero time-average. Furthermore, \citet{Weingartner2021} argued that the thermal flipping between the systematic torque relative to the angular momentum does not average in time to zero, which implies that thermal trapping is not prevalent. Nevertheless, these authors also showed that the DG mechanism is not capable of achieving the alignment of large grains. Instead, \citet{DM1976} proposed that radiative alignment torques (RAT) can rotate grains suprathermally, which was confirmed much later by \citet{DW1996,DW1997} using simulations. A simple analytical model (helical grain) by \citet{LH2007}, which was followed by a more complex set of grain shapes by \citet{Herranen2019}, resolved the mysterious dynamics of dust grains subjected to radiation and demonstrated that the alignment efficiency of large dust grains can explain observations. Thus, RATs is currently the leading theory describing grain alignment \citep{Andersson2015}.

The RAT theory describes the grain alignment as follows. Starlight radiation transfers angular momentum to the dust grains in the ISM. Dust grains absorb more radiation along one of their axes as a result of the asymmetric geometry nature of the dust grains. The dust grains end up spinning along the axis of greatest moment of inertia (i.e., the minor-axis of dust grain). As most dust grains are paramagnetic, they acquire a magnetic moment experiencing the fast Larmor precession. After a relaxation time, dust grains will spin along the local B-field \citep{Andersson2015,HL2016}. The final configuration is given by the long-axis of the dust grains to be perpendicular to the orientation of the local B-field. Thus, the absorptive polarization has a position angle (PA) of polarization parallel to the local B-field, while the emissive polarization has a PA of polarization perpendicular to the local B-field. This mechanism is known as B-RATs, and it is the most common tool for interpreting polarimetric observations tracing absorptive/emissive polarization. Indeed, the induced polarization from grain alignment by RATs has been extensively used to interpret observations from the ISM in the Milky Way \citep[][for a review]{Andersson2015} and galaxies \citep{ELR2018,Jones2019,ELR2020,Jones2020,ELR2021a,ELR2021b,Borlaff2021,Pattle2021b,SALSAIV,SALSAV,ELR2023}. However, when the B-fields are weak and/or the radiation is strong, the minor-axis of grains can be aligned with the radiation field instead of B-fields. This alignment is called k-RAT and may has been observed in the Orion molecular cloud \citep{Pattle2021}.

In addition, \citet{Gold1952} proposed that dust grains can be aligned by stochastic mechanical alignment of thermal rotation. However, this mechanism is found to be efficient for nanoparticles \citep{HT2019} and inefficient compared to mechanical force (or METs, \citealt{LH2007b,LH2021}). The METs fundamental is basically the same as RATs, by replacing a photon beam by a gas flow, and numerically shown to be efficient in aligning grains interacting with both subsonic and supersonic incident gas flow.   

In this paper, we aim to characterize the dust grain alignment mechanisms across the spiral galaxy M\,51. M\,51 is the most studied face-on spiral galaxy using radio \citep{Fletcher2011,Kierdorf2020} and FIR \citep{Jones2020,Borlaff2021} polarimetric observations. These studies have shown that M\,51 has an ordered spiral B-field morphology closely following the spiral arms of the galaxy in both radio and FIR polarization, tracing B-fields in the diffuse and dense ISM, respectively. However, \citet{Borlaff2021,SALSAV} and \citet{Surgent2023} found that the dense and diffuse ISM do not necessarily trace the same B-field morphology; differences in the B-fields are found in the outskirts (R$\ge7$ kpc) of the galaxy. In addition, the FIR polarization fraction is higher in the inter-arms than in the arms, and there is a decrease in polarization fraction with column density within each component. 

Although a unique view of the B-field in the multi-phase ISM has been provided, the effect of the dust grain alignment mechanisms in the measured thermal polarization as a function of the gas density, B-field strength, gas and dust temperature of the galaxy has not been studied in detail. For the first time, our work presents a two-dimensional study of the dust grain alignment mechanism in an external galaxy. We describe in Section \ref{sec:AligMech} the specifics of the several dust grains alignment mechanisms. The results and discussions are described in Section \ref{sec:DIS}. Our main conclusions are summarized in Section \ref{sec:CON}.


\section{Timescales for external alignment of dust grains} \label{sec:AligMech}

\begin{figure*}[ht!]
\includegraphics[scale=0.66]{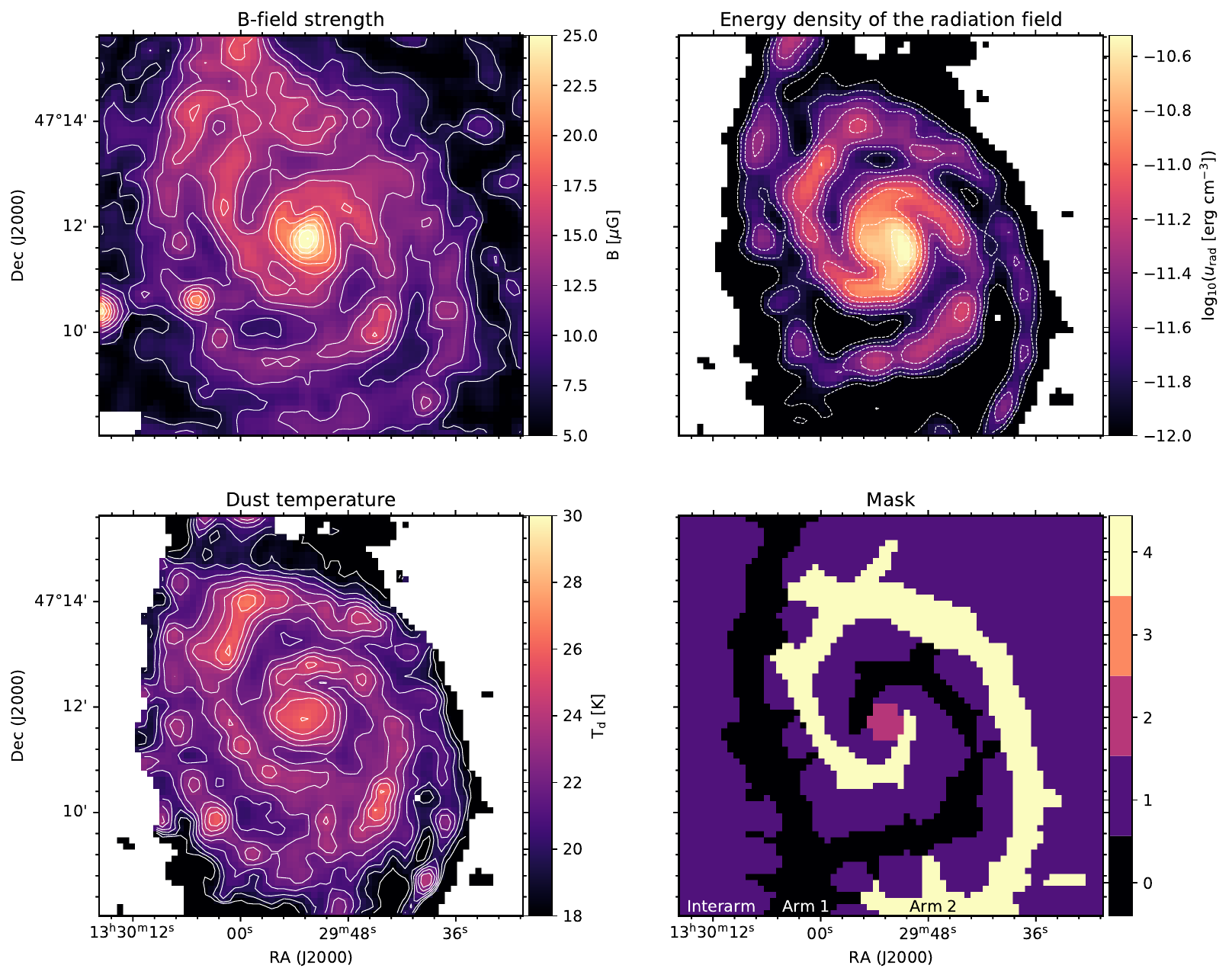}
\caption{M\,51's maps of the physical parameters to estimate the timescales of the dust grain alignment mechanisms. B-field strength ($B$, top left) by \citet{Fletcher2011} with contours starting at $7~\mu$G and increasing in steps of $2~\mu$G, 
energy density of the radiation field ($u_{rad}$, top right) as estimated by Eq. \ref{eq:urad} with contours starting at $\log_{10}(u_{\rm{rad}}~[\rm{erg~cm}^{-3}]) = -12$  and increasing in steps of 0.2, 
dust temperature as estimated in Section \ref{subsec:taular} ($T_{d}$, botton left) with contours starting at $18$ K and increasing in steps of $1$ K,
and the mask (bottom right) of the arms and interarms as estimated by \citet{Borlaff2021}.
 \label{fig:fig1}}
\end{figure*}

As discussed in the Introduction, dust grains are aligned either with B-fields, or radiation field, or the gas flow. To estimate which alignment mechanism plays a dominant role across the galaxy, we estimate the alignment timescales among B-RATs, k-RATs and METs, and compare them to the gas damping processes. We follow a similar approach as that presented by \citet{Draine1996,Tazaki2017,Pattle2021}. The typical range of dust grain sizes in the diffuse ISM is $0.005-0.3$ \um\ \citep{MRN1977,DL2007}. Hereafter, we take the dust grains sizes to be in the range of $a= 0.001 - 10~\mu$m to ensure we study a wider parameter space. Table \ref{tab:table1} summarizes the main equations and results from this section. 

\subsection{Larmor timescale (B-RAT)}\label{subsec:taular}
The condition for magnetically aligned dust grains (B-RAT) is such that the timescale for the dust grains precession around the B-field orientation must be shorter than other precession timescales. The Larmor timescale, $\tau_{\rm{Lar}}$, provides the timescale of  dust grains precessing around the B-field, which is given by

\begin{equation}
\tau_{\rm{Lar}} = \frac{2 \pi I_{\parallel} \Omega}{|\mu_{\rm{Bar}}| B}
\end{equation}
\noindent
\citep{HL2016}, where $I_{\|} = \frac{8\pi}{15}\rho s a^{5}$ is the greatest moment of inertia of the dust grain, $\rho$ is the dust grain density, $s = \frac{c}{a}$ is the axial ratio of oblate grains, $c$ is the length of the semi-minor axis and $a$ is the lengths of the semi-major axes (i.e., grain size), $\Omega$ is the angular velocity of the grain, $B$ is the magnetic field strength, and $|\mu_{\rm{Bar}}|$ is the grain magnetic moment given by

\begin{equation}
    |\mu_{\rm{Bar}}| = \frac{\chi(0) \omega V}{|\gamma_{\rm{e}}|}
\end{equation}
\noindent 
where $\omega = 0.47\Omega$ is the angular frequency of the grain, $V=\frac{4\pi}{3} s a^{3}$ is the dust grain volume, $|\gamma_{\rm{e}}| = \frac{e}{m_{\rm{e}} c}$ is the gyromagnetic ratio of an electron, and the susceptibility is given by Curie's law

\begin{equation}
    \chi(0) = \frac{n_{\rm{p}} \mu^{2}}{3 k_{\rm{B}} T_{\rm{d}}}
\end{equation}
\noindent
where $n_{\rm{p}}$ is the atomic density of the material, $\mu=p \mu_{B}$ is the effective moment per iron atom with $p=\gamma_{\rm{e}} \sqrt{J(J+1)} \sim 5.5$ for silicate grains and $\mu_{B}$ is the Bohr magneton, $k_{\rm{B}}$ is the Bolztman constant, and $T_{\rm{d}}$ is the dust temperature. 

Therefore, the Larmor timescale for a paramagnetic dust grains can be written, in practical units, as

\begin{equation}
\tau_{\rm{Lar,PM}} = 0.03 \rho_{3} a_{-5}^{2} n_{23}^{-1}f_{\rm{p},7}^{-1}p_{5.5}^{-2} T_{\rm{d},15} B_{5}^{-1}~~\rm{yr}
\end{equation}

\noindent
where $\rho_{3} = \rho/3$ g cm$^{-3}$, $a_{-5} = a/10^{-5}$ cm, $n_{23} = n/10^{23}$ cm$^{-3}$ and $n$ is the atomic density of the material, $f_{\rm{p},7} = 1/7$ is the fraction of paramagnetic atoms in a silicate grain with the structure MgFeSiO$_{4}$ \citep{HLM2014}, $p_{5.5}=p/5.5$, $T_{\rm{d},15} = T_{\rm{d}}/15$ K, and $B_{5} = B/5~\mu$G. The mass of a silicate dust grain with a structure MgFeSiO$_{4}$ is $m_{\rm{sil}}= 2.86\times10^{-22}$ g and mass density $\rho_{\rm{sil}}=3.81$ g cm$^{-3}$, yields an atomic density of $n_{\rm{sil}}=1.33\times10^{22}$ cm$^{-3}$. 

To compute the dust temperature, we binned the $70-350$ \um~PACS/SPIRE/\textit{Herschel} observations \citep{Pilbratt2010, Poglitsch2010, Griffin2010} to a pixel scale, $3\farcs2$, equal to the pixel scale of the $154$ \um\ HAWC+/SOFIA observations of M\,51 \citep{Borlaff2021}. Then, we extracted the intensity values of each pixel associated with the same part of the sky at each wavelength. Finally, for every pixel we fit a modified blackbody function with a constant dust emissivity index of $\beta = 2.0$. \citet{MC2012} used a similar approach than that presented here but allowing $\beta$ ranging from $1$ to $3$, with the best fit at $\beta=2.0\pm0.4$ across the whole galaxy. Our estimated dust temperatures (Figure \ref{fig:fig1}) across the galaxy range from $18-26$ K, in agreement with \citet{MC2012}. 

Alternatively, \citet{Draine1996} computed that the  dust temperature can also be estimated for silicates using the relation

\begin{equation}
T_{\rm{d}} = 16.4 \left( \frac{u_{\rm{rad}}}{u_{\rm{ISRF}}} \right)^{1/6}~\mbox{K}
\label{eq:Td}
\end{equation}
\noindent
where $u_{\rm{rad}}$ is the energy density of the radiation field, $u_{\rm{ISRF}}$ is the energy density of the standard interstellar radiation field (ISRF), such as $u_{\rm{ISRF}} = 8.64\times10^{-13}$ erg cm$^{-3}$. 

For optically thin FIR emission, the energy density of the radiation field can be estimated as 

\begin{equation}
u_{\rm{rad}} = 3\times10^{-9} \left( \frac{\Sigma_{\mbox{g}}}{\mbox{g cm}^{-2}} \right)^{N}~\mbox{erg cm$^{-3}$}
\label{eq:urad}
\end{equation}
\noindent
\citep{Lacki2010} where $N=1.4$ is the exponent of the Kennicutt--Schmidt relation \citep{Kennicutt1998}.
$\Sigma_{\rm{g}}$ is the gas surface density, which we estimate as $\Sigma_{\rm{g}} = \mu\,m_{\rm{H}}\, N_{\rm{H2+HI}}$. $\mu$ is the mean molecular weight per hydrogen atom $\mu = 2.8$, $m_{\rm{H}}$ is the hydrogen mass and $N_{\rm{H2+HI}}$ is the density of the gas column. We take the $N_{\rm{H2+HI}}$ of M~51 from \citet{Borlaff2021}. These authors estimated the column density using the HI neutral gas and $^{12}$CO(1-0) molecular gas. Figure \ref{fig:fig1} shows the energy density of the radiation field, $u_{\rm{rad}}$. We estimate $T_{\rm{d}}$ using Eqs.\,\ref{eq:Td} and \ref{eq:urad} and compare it with the dust temperature estimated using the \textit{Herschel} data. We find that on average, both dust temperatures agreed within $15$\%. 

For completeness with the literature on the applicability of RATs in galaxies, \citet{Hoang2021b} studied the effect of redshift with dust grain alignment properties in galaxies and used the relationship between temperature and redshift $z$ of $T_{\rm{d}} = (34.6\pm0.3) + (3.94\pm0.26)(z-2)$ K by \citet{Bouwens2020}. This relationship was derived from photometric data points of entire galaxies spanning $z=0-10 $. Using $z=0.00131$ \citep{Chen2018} for M\,51, we estimate a median of $T_{\rm{d}} = 26.7\pm0.8$ K, which is slightly higher than the median value of $21.8\pm2.1$ K from our estimated dust temperature using \textit{Herschel} data.

In the following, we take the dust temperature map (Figure \ref{fig:fig1}) estimated using the \textit{Herschel} data as this map provides the characteristic temperature of the spectral energy distribution of $70-350~\mu$m of dust emission using empirical data of M\,51, instead of the proxy of dust temperatures by \citet{Draine1996} and \citet{Hoang2021b}.

Using the dust temperature maps shown in Figure \ref{fig:fig1}, we estimate a median $\chi(0) \sim 4\pm2\times10^{-3}$ across the galaxy. Note that the magnetic susceptibility has been estimated to be in the range of $4.2\times10^{-5} - 4.2\times10^{-3}$ for paramagnetic dust grains \citep{Draine1996}. To estimate the uncertainty of $\tau_{\rm{Lar,PM}}$, we vary the range of $\chi(0)$ by a factor of 10.

We take the total B-field strength map of M\,51 calculated using the radio polarimetric observations by \citet{Fletcher2011}. The B-field strength was estimated by assuming equipartition between the energy densities of magnetic fields and total cosmic rays, a constant ratio of cosmic ray protons and electrons of $100$, and a constant path-length through the synchrotron-emitting disc of $1$~kpc.
Figure \ref{fig:fig1} shows the map of the total B-field strength. 

Finally, we show the the Larmor timescale as a function of the dust grain sizes in the arms and interarms in Figure \ref{fig:fig2}. The uncertainties were estimated as follows. First, the galaxy was separated in arms 1 and 2, and interarm using the mask (Fig. \ref{fig:fig1}) produced by  \citet{Borlaff2021}. The mask was computed using the HI emission line, which emission arises from the arms. The nucleus (i.e., central two beams in diameter) of the galaxy has been masked for all of the analysis presented in this work. For each galaxy's component, the median timescale and their standard deviation per dust grain size were estimated. Second, the Larmor timescale for the ranges of the magnetic susceptibility was estimated. The final uncertainties (the shadowed region in Fig. \ref{fig:fig2}) were estimated as the sum in quadrature of these two uncertainties. The wide range of magnetic susceptibility dominates the uncertainties of the Larmor timescale. Results are discussed in Section \ref{subsec:AlignMech}.

\subsection{Radiative precession timescale (k-RAT)}\label{subsec:taurad}

Assuming that the dust grains rotate at the thermal angular velocity \citep{LH2007}, the radiative precession timescale, $\tau_{\rm{k-RAT}}$, is given by 

\begin{eqnarray}\label{eq:tauKRAT}
\tau_{\rm{k-RAT}} &=&  \frac{2 \pi I_{\parallel} \Omega_{\rm{T}}}{\gamma u_{\rm{rad}} \lambda a_{\rm{eff}}^{2} Q_{\rm{e3}}} \\
& =& 80.3 \rho_{3}^{1/2} s_{0.5}^{-1/6} a_{-5}^{1/2} T_{\rm{g,10}}^{1/2} \gamma^{-1} U^{-1}  \lambda_{1.2}^{-1} Q_{\rm{e3,-2}}^{-1}~\mbox{yr}
\end{eqnarray}
\noindent
\citep{Draine1996,Tazaki2017,TH2022}, where $\Omega_{\rm{T}} = \left(\frac{2k_{\rm{B}}T_{\rm{g}}}{I_{}\parallel}\right)^{1/2}$ is the thermal angular velocity of the gas with $T_{\rm{g}}$ as the gas temperature, $\lambda$ is the mean wavelength of the incident radiation, $\gamma$ is the radiation field anisotropy, $a_{\rm{eff}}=s^{1/3}a$ is the effective grains size, and  $Q_{\rm{e3}}$ is the third component of RATs that induces the grain precession around the radiation direction \citep{LH2007}. In practical units, $\rho_{3}$ and $a_{-5}$ are taken as above, $s_{0.5}=s/0.5$, $T_{\rm{g},10} = T_{\rm{g}}/10$ K, $U=\frac{u_{\rm{rad}}}{u_{\rm{ISRF}}}$, $\lambda_{1.2} = \lambda/1.2$ \um, and $Q_{\rm{e3,-2}} = Q_{\rm{e3}}/10^{-2}$.

\begin{figure*}[ht!]
\includegraphics[scale=0.68]{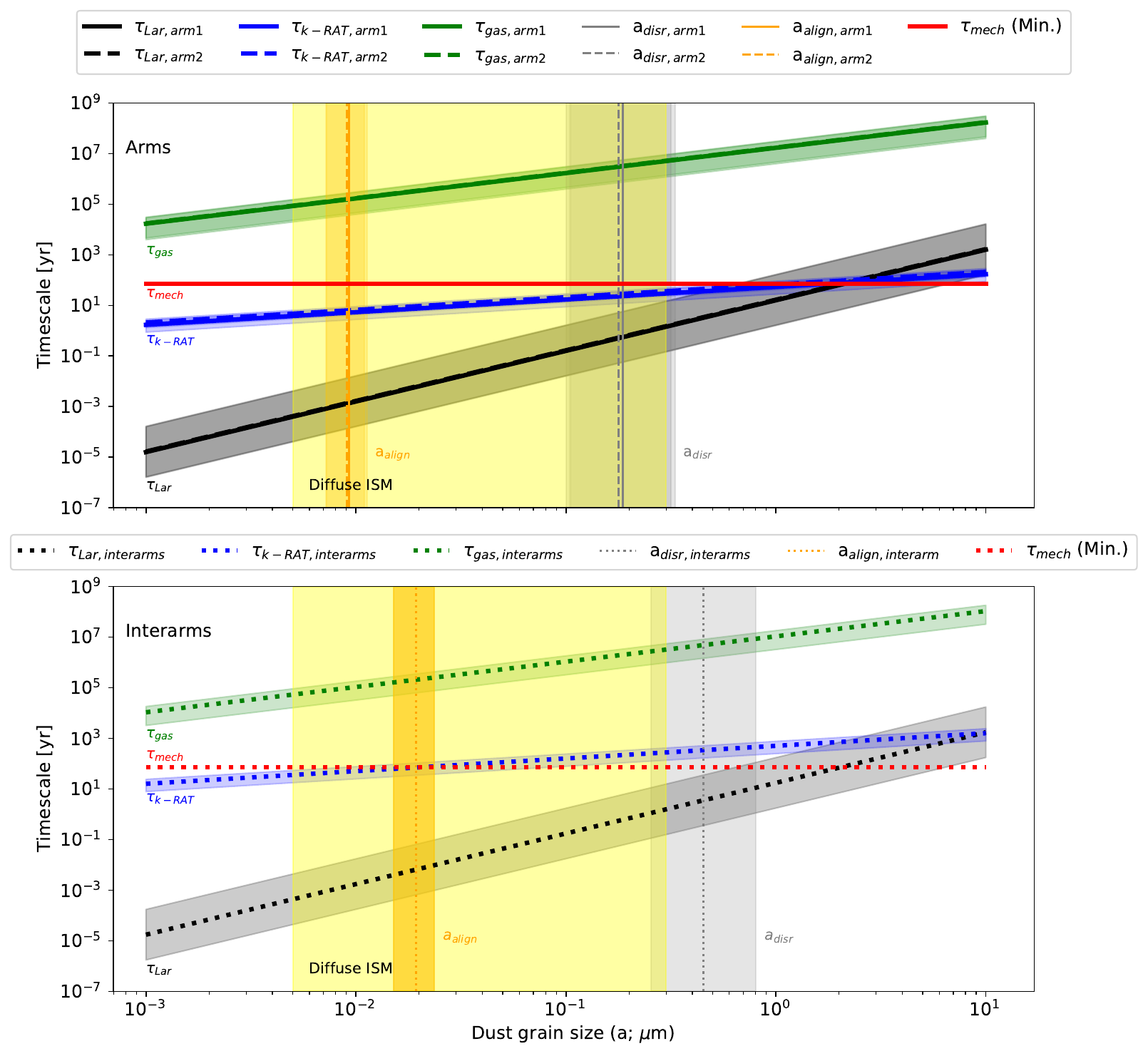}
\caption{Timescales of the several grain alignment mechanisms as a function of the dust grain size, $a$, for the arms (top) and interams (bottom). The timescales of Larmor (B-RAT; $\tau_{\rm{Lar}}$) (black), radiative precession (k-RAT, $\tau_{\rm{k-RAT}}$, blue),  gas damping (randomization; $\tau_{\rm{gas}}$, green), and mechanical alignment ($\tau_{\rm{mech}}$, red) are shown. The timescales for arm~1 (solid line), arm~2 (dashed line), and interarm (dotted line) are shown. The typical dust grain sizes ($0.005-0.2$ \um) in the diffuse ISM are shown as a yellow shaded region. The dust grain sizes at which the dust grains become aligned, $a_{\rm{align}}$ (orange), and disrupted by the incident radiation, $a_{\rm{disp}}$ (grey), are shown.
\label{fig:fig2}}
\end{figure*}

\begin{deluxetable*}{llccc}[ht!]
\tablecaption{Timescales of grain alignment mechanisms and dust grain sizes of the dominant alignment mechanism.\label{tab:table1}}
\tablewidth{0pt}
\tablehead{\colhead{Mechanism}	&	\colhead{Definition}	& 	\colhead{Physical structure}	& \colhead{Dust grain size regime}	\\ 
	&	&	&	\colhead{(\um)}		
}
\startdata
B-RAT	&	$\tau_{\rm{Lar,PM}}$	$\propto a_{-5}^{2}T_{\rm{d},15}B_{5}^{-1}$	& Arms	&	$0.009-0.182$		\\
		&					&	Interarms	&	$0.019-0.452$		\\
k-RAT	&	$\tau_{\rm{k-RAT}} \propto a_{-5}^{1/2}T_{g}^{1/2}u_{rad}^{-1}\gamma^{-1} \propto a_{-5}^{1/2}T_{g,10}^{1/2}\Sigma_{g}^{-1.4}\gamma^{-1}$ & Arms	&	not satisfied	 \\
		&				&	Interarms		&	not satisfied	\\
Randomization	&	$\tau_{\rm{gas}} \propto a_{-5}T_{\rm{g,10}}^{-1/2}n_{\rm{H,3}}^{-1} \propto a_{-5}T_{\rm{g,10}}^{-1/2}\Sigma_{\rm{g}}^{-1}$	&	Arms 	&	not satisfied	 \\
		&				&	Interarms		&	not satisfied	 \\
v-MAT   &   $\tau_{\rm{mech}}$  $\ge 72$ & Arms    &   not satisfied     \\
    &   &   Interarms   &   not satisfied    \\
Min. disruption   &   $a_{\rm{disr}} \propto \gamma^{-1/2}n_{H,3}^{1/2}u_{rad}^{-1/2}T_{gas}^{1/2} \propto \gamma^{-1/2} \Sigma_{g}^{-0.2}T_{gas}^{1/2}$   &   Arms &   $\ge0.182$      \\
        &              &   Interarms   &   $\ge0.452$   \\
Min. alignment   &   $a_{\rm{align}} \propto \gamma_{0.1}^{-2/7}n_{H,3}^{2/7}u_{rad}^{-2/7}T_{gas}^{2/7} \propto \gamma_{0.1}^{-2/7} \Sigma_{g}^{-3.5}T_{gas}^{2/7}$   &   Arms    &   $\ge0.009$    \\
        &               &   Interarms   &   $\ge0.019$    \\
\enddata
\end{deluxetable*}


The heating source for the dust grains in M\,51 is thought to arise from the evolved stellar population heating mechanism, although heating by ongoing star formation and/or by active nuclei cannot be ruled out \citep{MC2012}. We took $\lambda = 1.2$ \um\ ($\lambda_{1.2} = 1$), as the typical wavelength peak of the radiation of the ISRF in our Galaxy \citep{Hoang2020}. Radiation field anisotropy is in the range of $0 < \gamma \le 1$, with $\gamma = 0.1$ in the diffuse ISM \citep{Draine1996}, $\gamma=0.7$ for molecular clouds \citep{DW1996}, and $\gamma \le 1$ near a protostar \citep{Tazaki2017}. However, the exact number to be used under specific physical conditions is very complex. \citet{Bethell2007} show that the description between anisotropy and density structure (i.e., clumpiness) in the ISM is still unclear, after using highly inhomogeneous simulations of compressible magnetohydrodynamics (MHD) turbulence. 

To compute $\gamma$ on a pixel-by-pixel basis, we use the map of the star formation rate (SFR) of the galaxy as a proxy of the radiation received by the dust grains across the galaxy. Specifically, we used a two-wavelength hybrid tracer to estimate the SFR surface density map $\Sigma_{\rm{SFR}}$ as

\begin{equation}
\Sigma_{\rm{SFR}} = 8.85\times10^{-2} I_{\rm{FUV}} + 3.02 \times 10^{-3} I_{\rm{W4}}~~~\rm{M_{\odot}~yr^{-1}~kpc^{-2}}
\end{equation}
\noindent
where $I_{\rm{FUV}}$ is the intensity at far-ultraviolet (FUV) wavelengths using \textit{GALEX} in units of MJy sr$^{-1}$, and $I_{\rm{W4}}$ is the intensity at W4 using \textit{WISE} in units of MJy sr$^{-1}$. The GALEX and WISE observations were taken from \citet{Leroy2019}. This approach corrects the dust attenuation in the UV wavelength regime using IR data. 

We assume that dust in areas of high SFR will have a maximum gamma, while areas with a minimum SFR will have a minimum $\gamma$. Taking this approach, we rescale the SFR map with a range of $\gamma$ values in every pixel of the galaxy. We take a minimum value of $\gamma=0.1$ for the minimum of the SFR, and a maximum value of $\gamma=0.8$ for the maximum of the SFR.

The temperature of the molecular gas in the giant molecular clouds along the arms of M\,51 has been measured to be $\sim20$ K \citep{Schinnerer2010} using high-resolution interferometric observations of $^{12}$CO(2-1), $^{12}$CO(1-0), and $^{12}$C$^{18}$O. Further interferometric and single-dish telescopes, as part of the Plateu de Bure Interferometer Arcsecond Whirpool Survey (PAWS), measured an extended diffuse ISM with a temperature $<200$ K and a scale height of $200$ pc \citep{Pety2013}, which accounts for $\sim50$\% of the total $^{12}$CO(1-0) emission. We take the gas temperature in the arms to be similar to the dust temperature, $T_{\rm{g,arms}} \sim T_{\rm{d}}$. However, the gas temperature in the diffuse ISM of M\,51 is assumed to be equal to the upper-limit of the diffuse and cold phase of the ISM, $T_{\rm{g,interarms}} = 200$ K.



Figure \ref{fig:fig2} shows the radiative precession timescale as a function of the dust grain sizes and galaxy's components. For each galaxy's component, the median timescales (lines) and their standard deviation (shadowed region) per dust grain size were estimated. Results are discussed in Section \ref{subsec:AlignMech}.

\subsection{Gas damping timescale (Randomization)}\label{subsec:taugas}
The timescale of dust grains randomization by gas collisions, $\tau_{gas}$, is estimated such as

\begin{equation}
\tau_{\rm{gas}} = 6.26 \times 10^{3} \rho_{3} s_{0.5}^{-2/3} a_{-5} \Gamma_{||}^{-1} T_{\rm{g,10}}^{-1/2} n_{\rm{H,3}}^{-1}  ~\mbox{yr}
\label{eq:taugas}
\end{equation}
\noindent
\citep[i.e.,][]{Draine1996,HL2016}, where $\rho_{3}$, $s_{0.5}$, $a_{-5}$, $T_{\rm{g,10}}$ are taken as above. $\Gamma_{||} \sim 1$ is a factor of order unity describing the dust grain geometry, and $n_{\rm{H}}$ is the gas volume density in units of cm$^{-3}$, normalized as $n_{\rm{H,3}} = n_{\rm{H}}/10^{3}$ cm$^{-3}$

The gas volume density is estimated such as $n_{\rm{H}} = N_{\rm{H2+HI}}/h$, where $h$ is the vertical height of the galactic disk.  We take that most of the dust is colocated in the midplane of the galaxy, with a vertical height of $h = 200$ pc  for M\,51 \citep{Pety2013,MA2023}. 

Figure \ref{fig:fig2} shows the gas damping timescale as a function of the dust grain sizes galaxy's components. For each galaxy's component, the median timescales (lines) and their standard deviation (shadowed region) per dust grain size were estimated. Results are discussed in Section \ref{subsec:AlignMech}.

\subsection{Mechanical alignment timescale (v-MAT)}\label{subsec:taumech}

The velocity vector of the gas and dust grain drift can produce a preferential precession \citep[v-MAT;][]{LH2007}. When the gas and dust grain velocity difference is subsonic and the precession time around the gas flow, $\tau_{\rm{mech}}$, is smaller than the Larmor precession timescale, the v-MAT alignment mechanism can dominate, i.e., $\tau_{\rm{mech}} < \tau_{\rm{Lar}}$. In addition, this mechanism requires dust grains with large helicity acquired through coagulation \citep[i.e.][]{Brauer2008}. The mechanical alignment timescale is given by 

\begin{equation}
\tau_{\rm{mech}} = 36\left( \frac{c_{s}^{2}}{\Delta v} \right )^{2} \left( \frac{\omega}{\omega_{\rm{th}}} \right) s_{0.5}^{2} \frac{1}{\sin 2\theta}~\mbox{yr}
\label{eq:taumech}
\end{equation}
\citet{LH2021}, where $s_{0.5}$ is taken as above. $c_{s}$ is the gas sound speed, $\Delta v$ is the gas and dust grain velocity difference, $\omega$ is the dust grain angular velocity, $\omega_{\rm{th}}$ is the thermal angular velocity, and $\theta$ is the angle between the major axis of inertia of the dust grain and the direction of the radiation. 

We take $\sin {2\theta} = 0.5$ \citep{LH2021}, and the condition for v-MAT that $\Delta v < c_{s}$, thus $\tau_{\rm{mech}} > 72 (\omega/\omega_{\rm{th}})$ yr. The condition for v-MAT is that $\tau_{\rm{mech}} < \tau_{\rm{Lar}}$,  yields $\tau_{\rm{mech}} < 72$ yr and consequentially $\omega / \omega_{\rm{th}} < 1$. This result implies that the dust grains are rotating subthermally, which is physically implausible. We take the minimum condition for v-MAT to be $\tau_{\rm{mech}} \ge 72$ yr (Figure \ref{fig:fig2}). Results are discussed in Section \ref{subsec:AlignMech}.

\subsection{Grain alignment and disruption (RAT-D)}\label{ssubsec:RATd}

Eq. \ref{eq:tauKRAT} demonstrates that the efficiency of the k-RAT mechanism (radiative precession), denoted $\tau_{\rm{k-RAT}}$, is proportional to $a_{-5}^{1/2}u_{\rm{rad}}^{-1}T_{\rm{g,10}}^{1/2}$. This indicates that for large dust grains in strong radiation fields assotiated with star-forming regions in the spiral arms, the k-RAT mechanism becomes more efficient than other alignment mechanisms (Fig. \ref{fig:fig2}). However, \citet{Hoang2019,Hoang2019b} have shown that strong incident radiation fields can cause the breakdown of large dust grains. This phenomenon, called RAT-D, alters the population of the dust grain population by increasing the abundance of small dust grains compared to large dust grains. RAT-D is based on the concept of a critical size of the dust grain, denoted $a_{\rm{disr}}$, at which the tensile strength of the dust grain decreases in comparison to incident radiation. When $a \ge a_{\rm{disr}}$, the dust grains are disrupted. Estimation of the size of the dust grain, $a_{\rm{disr}}$, at which rotational disruption occurs, can be taken as follows

\begin{eqnarray}
a_{\rm{disr}} & = 0.22\rho_{3}^{-1/4}\gamma^{-1/2} n_{\rm{H,3}}^{1/2}  U^{-1/2} T_{\rm{g,10}}^{1/4} \\ \nonumber
& \lambda_{0.5}S_{\rm{max},7}^{1/4}(1+F_{\rm{IR}})^{1/2}~\mbox{\um} 
\label{eq:adisp}
\end{eqnarray}
\citep{Hoang2021} where $\rho_{3}$, $\gamma$, $n_{\rm{H,3}}$, $U$, $T_{\rm{g,10}}$ are defined as above. $\lambda_{0.5} = \lambda/1.5$ \um,  $S_{\rm{max},7} = S_{\rm{max}}/10^{7}$ erg cm$^{-3}$ with $S_{\rm{max}}$ as the tensile strength determined by the internal structure of dust grains \citep{Hoang2021}, and $F_{\rm{IR}}$ is a dimensionless parameter that describes the ratio of the IR damping to the collisional dumping rate, which is

\begin{equation}
F_{\rm{IR}} = 0.4 U^{2/3} a_{-5}^{-1} n_{\rm{H,30}}^{-1}T_{\rm{g,100}}^{-1/2}
\label{eq:FIR}    
\end{equation}
where $n_{\rm{H,30}} = n_{\rm{H}}/30$ cm$^{-3}$, and $T_{\rm{g,100}} = T_{\rm{g}}/100$ K.

In addition, it is important to note that not all grains can be aligned using RATs. Only grains that are rotated at an angular velocity that is more than three times the thermal angular velocity are able to be effectively aligned. The critical grain size is called $a_{\rm align}$, and the dust grains align with $a \ge a_{\rm{align}}$. Thus, the ISM has aligned dust grains within the range of sizes given by $[a_{\rm{align}},a_{\rm{disr}}]$. 

The minimum size of aligned dust grains can be estimated as

\begin{eqnarray}
a_{\rm{align}} & = 0.055 \rho_{3}^{-1/7} \gamma_{0.1}^{-2/7} n_{\rm{H},3}^{2/7} U^{-2/7} T_{\rm{gas},10}^{2/7} \\ \nonumber
& \lambda_{1.2}^{4/7} (1+F_{\rm{IR}})^{2/7}~\mbox{\um}
\label{eq:amin}
\end{eqnarray}
\noindent
where $\lambda_{0.1} = \lambda/0.1$ \um. 

The efficiency of RAT-D is mainly determined by the tensile strength $S_{\rm max}$ \citep[e.g.,][]{Hoang2019}. However, this quantity is unclear in the ISM. The uncertainty in $S_{\rm max}$ is influenced by multiple factors. For example, in a dense cloud, dust is expected to have a more complex composition, which could result in a decrease in $S_{\rm max}$. On the other hand, in a less dense medium, dust particles may be more compact, leading to an increase in $S_{\rm max}$. In order to compensate for the compactness of dust, we use a value of $S_{\rm max}=10^{7}\,\rm erg\,cm^{-3}$ in the arm and $S_{\rm max}=10^{9}\,\rm erg\,cm^{-3}$ in the interarms.

\begin{figure*}[ht!]
\includegraphics[width=\textwidth]{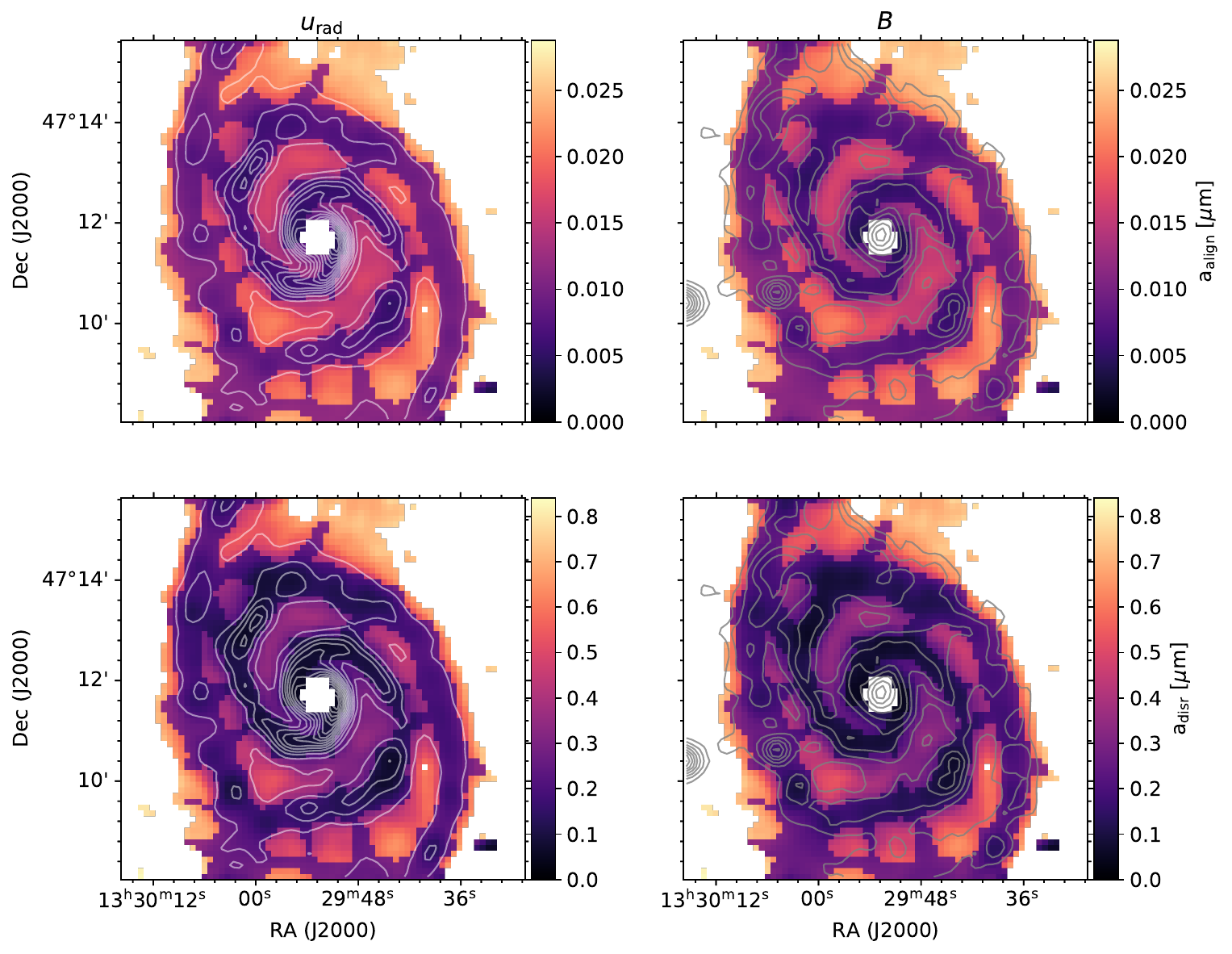}
\caption{Align (top) and disrupted (bottom) dust grain sizes across the disk of M\,51. The contours show the radiation field, $u_{\rm{rad}}$, (white) starting at $10^{-12}$ erg cm$^{-3}$ and increasing in steps of $2.07\times10^{-12}$ erg cm$^{-3}$, and the magnetic field strength, $B$, (grey) starting at $10$ $\mu$G and increasing in steps of $2$ $\mu$G.
\label{fig:fig3}}
\end{figure*}

The aligned and disrupted dust grain sizes in the disk of M\,51 are illustrated in Figure \ref{fig:fig3}, and the median values and standard deviation within the arms and interarms are shown in Figure \ref{fig:fig2}. It is evident that the arms contains smaller grains aligned than in the interarms, and that the dust grains are disrupted at smaller sizes than in the interarms. The median dust grain sizes denoted as [$a_{\rm{align}}$, $a_{\rm{disr}}$] are estimated to be $[0.009,0.182]$ \um\ and $[0.019,0.452]$ \um\ for the arms and interarms, respectively. No statistically significant differences were found between arms 1 and 2, we analyze both arms together in the subsequent sections. These estimations are discussed in Section \ref{subsec:Pem}.

For completeness, in addition to the mechanisms presented here, we estimate the timescale of the internal alignment by the Barnett relaxation \citep{Purcell1979} in Appendix \ref{sec:AligMechInternal}. We estimate that very large, i.e., mm-size, dust grains are required for this alignment mechanism to be comparable to the timescales of those presented here. Thus, perfect internal alignment is present in FIR polarimetric observations of galaxies.


\section{Discussion} \label{sec:DIS}

\subsection{Dominant dust grain alignment mechanism}\label{subsec:AlignMech}

The standard size of dust grains in the ISM is in the range of $0.05-0.3$ \um\ \citep{MRN1977,DL2007}, with the growth of dust grains occurring on small scales within molecular clouds and star-forming regions \citep[i.e.][]{Draine1990,Jenkins2009}. Within the size range of aligned dust grains, [$a_{\rm{align}}$, $a_{\rm{disr}}$], we estimate that the Larmor timescale (B-RAT) is always faster than radiative precession timescales (k-RAT), mechanical alignment ($\nu$-MAT), and gas dumping mechanism (randomization), i.e., $\tau_{\rm{Lar,PM}} < \tau_{\rm{k-RAT}} \le \tau_{\rm{mech}} << \tau_{\rm{gas}}$.  Although k-RAT and v-MAT mechanisms show smaller timescales than B-RAT at large dust grains ($a>2$ \um), incident radiation in both arms and interams is strong enough to disrupt large dust grains in the ISM of M\,51 (Section \ref{subsec:Pem}). 

The larger uncertainty in the B-RAT mechanism is given by the estimation of the magnetic susceptibility. The magnetic susceptibility increases with increasing magnetic inclusions in dust grains \citep{JS1967}, however, this is a complex parameter to quantify. \citet{LH2021,Herranen2021} performed a statistical analysis of the alignment efficiency, $\delta_{\rm{m}} = \tau_{\rm{ran}}/\tau_{\rm{Lar}}$, as a function of the magnetic inclusions (i.e., magnetic susceptibility), where $\tau_{\rm{ran}}$ is defined as the random alignment given by the addition of all timescales except the Larmor timescale. \citet{Herranen2021} found that even for low magnetic inclusions an alignment efficiency of $0.5$ can explain the polarization observations in the ISM by B-RAT. We estimate dust grains size ranges, [$a_{\rm{align}}$, $a_{\rm{disr}}$], of $[0.009,0.182]$ \um\ and $[0.019,0.452]$ \um\ for the arms and interarms. Within these ranges, $\tau_{\rm{Lar,PM}}$ is $2-4$ orders of magnitude smaller than $\tau_{\rm{k-RAT}}$ and $\tau_{\rm{mech}}$ in the arms and interarms, respectively (Fig. \ref{fig:fig2}). Despite the wide range of magnetic susceptibility, we conclude that the dominant dust grain alignment mechanism in M\,51 arises from magnetically aligned dust grains (B-RAT).

Furthermore, k-RAT has the closest timescale to B-RAT in the arms, but is still $\ge1$ order of magnitude less efficient than B-RAT. For k-RAT to dominate, the B-field should be $<0.6$ $\mu$G or the radiation field of the galaxy to be $>10^{2}$ times larger than the $u_{\rm{ISRF}}$ of the ISM. In the interarms, the mechanical alignment (v-MAT), $\tau_{\rm{mech}}$, is the closest timescale scale to B-RAT, but is still $\ge1$ order of magnitude less efficient than B-RAT. For v-MAT to dominate, the B-field should be $<0.6$ $\mu$G or the dust temperature to be $>10^{2}$ times larger than the sublimation temperatures, $\sim1500$ K, of silicates. Neither of these conditions  are compatible with typical measurements in spiral galaxies. Thus, we conclude that magnetically aligned dust grains may be the dominant dust grain alignment mechanism in spiral galaxies.

\subsection{Origin of polarized emission}\label{subsec:Pem}

Figure \ref{fig:fig3} shows the map of the minimum, $a_{\rm{align}}$, and maximum, $a_{\rm{disr}}$, dust grain sizes to satisfy $\tau_{\rm{Lar}} < \tau_{\rm{k-RAT}}$ across the galaxy. We find that the arms are dominated by small dust grains (bluer colors), while the interarms are dominated by large dust grains (redder colors). For the arms, the incident radiation from star-forming regions is disrupting the large dust grains and have enough strength to align the small dust grains. Thus, arms have a dust grain size population enhanced by small dust grains up to an estimated $a_{\rm{disp,arms}} \leq 0.182$ \um. For the interarms, the incident radiation is not strong enough to align small dust grains, $a_{\rm{align,interarms}} \geq 0.019$ \um, but is strong enough to generate a torque to align large dust grains up to $a_{\rm{disr,interams}} \leq 0.452$ \um. Thus, interarms have a dust grains size population missing small dust grains in the $0.009-0.019$ \um\ regime and have an additional population of larger dust grains in the $0.182-0.452$ \um\ range compared to the arms.

Using RAT and RAT-D theory \citep[i.e.,][]{Hoang2021b,TH2022}, \citet{Lee2020} showed that the dust grain alignment population affects the polarization fraction. Specifically, the polarization fraction decreases as a result of the enhancement of small dust grains as the incident radiation field increases. This effect is due to the disruption of large dust grains, which produce polarized emission more efficiently than small dust grains. These authors found that this effect is more clearly visible in dense molecular clouds radiated by nearby stars. The diffuse ISM also experiences this effect, but is less severe due to the lower radiation field compared to the tensile strength of the dust grains. \citet{Hoang2020} also suggested a decrease in the polarization fraction with increasing radiation strength (i.e., emission intensity, optical depth) due to the disturbance effect towards the cores of dense molecular clouds. 

For M\,51, the loss of large dust gains in the arms and the loss of small dust grain sizes in the interarms should produce a larger polarization fraction in the interarms than in the arms. Indeed, \citet{Borlaff2021} measured a higher FIR polarization fraction in the interarms, $P_{\rm{interarms}} = 4.2^{+0.3}_{-0.3}$\%, than in the arms, $P_{\rm{arms,1}} = 3.0^{+0.3}_{-0.3}$\% for arm~1 and $P_{\rm{arms,2}} = 3.5^{+0.4}_{-0.3}$\% for arm 2. Under the ideal assumption of no turbulent or geometrical effects, we estimate that the difference in polarization fraction between arms and interarms in M\,51 found by \citet{Borlaff2021} can also be affected from the enhancement of small dust grains sizes in the arms as an effect of the grain alignment disruption (RAT-D).

In addition, Figure \ref{fig:fig3} shows a comparison of the dust grain size across the galaxy with the energy density of the radiation field, $u_{\rm{rad}}$ (left panel), and B-field strength, $B$, (right panel). As expected by RATs, the disrupted dust grain sizes are spatially correlated with the energy density of the radiation field. The regions with smaller dust grains correspond to regions with higher $u_{\rm{rad}}$. The interarms show the highest dust grain sizes for the lowest B-field strengths. The arms show to have lower $a_{\rm{disr}}$ in the regions of strongest B-field strength. \citet{Borlaff2021} found that the polarization fraction decreases as a function of the total intensity (and column density) within the arms and interarms. Note that the dominant grain alignment is B-RAT, so differences in grain alignment mechanisms are negligible here. The depolarization effects measured in the polarization fraction versus the total intensity (or column density) in galaxies \citep{ELR2018,ELR2021b, Borlaff2021,SALSAIV} require explanation by additional physical mechanisms related to the B field structure along the LOS and/or turbulence fields, rather than only by the loss of dust grain alignment mechanisms. For example, changes in the depolarization rate with column density and dust temperature are found on regions with strong star formation across the disks of spiral galaxies \citep{SALSAIV}. This effect can be produced by an enhancement of the disruption effect in regions intense SFR due to an increase in the radiation field, in combination with an increment of the turbulent and/or tangled B-fields, which average out the net B-field orientation within the beam of the observations, resulting in a depolarization effect. Depolarization effects in the diffuse ISM of the MW has been measured by \textit{ Planck} \citep{PlanckXIX_2015,PlanckXX_2015}, where the polarization fraction decreases due to projection effects, tangling and variation of B-fields along the LOS.


\section{Conclusions} \label{sec:CON}

We have presented the first empirical study of the several dust grain alignment mechanisms in the ISM of the nearby spiral galaxy, M\,51. Our analyses show that magnetically aligned dust grains (B-RAT) is the dominant dust alignment mechanism across the galaxy. We estimated that other alignment mechanisms, randomization (gas dumping), radiative precession (k-RAT), and mechanical alignment (v-MAT), are negligible in M\,51. 

We estimated that the dust grain sizes change as a function of the galaxy's components. On the one hand, due to the higher energy density radiation field driven by star-forming regions in the arms than in the interarms, large dust grains in the arms are disrupted via the RAT-D mechanism. This result implies that the arms have an enhancement of small dust grain sizes. On the other hand, the low radiation field in the interarms is not able to align the smaller dust grain sizes. Specifically, we found the range of aligned dust grains to be [$a_{\rm{align}}$, $a_{\rm{disr}}$] = [$0.009$, $0.182$] \um\ and  [$0.019$, $0.452$] \um\ for the arms and interarms respectively. RAT-D \citep{Hoang2021} predicts a decrease of polarization fraction as the medium loses the large dust grain sizes. Thus, the lost of large dust gains sizes in the arms and the lost of small dust grain sizes in the interarms should produce a larger polarization fraction  in the interarms than in the arms. The difference in dust grain size population may be able to explain the  difference in polarization fractions between arms and interarms in M\,51 measured by \citep{Borlaff2021}. The polarization difference arises from the enhancement of small dust grains sizes in the arms as an effect of the grain alignment disruption.

Although RAT-D may explain the differences in polarization fraction between arms and interarms, it is required to further study its effect on the decrease of the polarization fraction with column density and gas dispersion within each of these components. We argue that the depolarization effects may be mainly due to intrinsic variations of the B-field structure in each galaxy's component and/or turbulence within the beam of the observations. Further MHD simulations of galaxies with the addition of dust grain alignment mechanisms are required to test their effect of the measured FIR polarimetric observations.


\begin{acknowledgments}
We thank Rainer Beck, Alejandro Borlaff, Susan Clark for their useful comments and discussions during this work. E.L.-R. is supported by the NASA/DLR Stratospheric Observatory for Infrared Astronomy (SOFIA) under the 08\_0012 Program.  SOFIA is jointly operated by the Universities Space Research Association,Inc.(USRA), under NASA contract NNA17BF53C, and the Deutsches SOFIA Institut (DSI) under DLR contract 50OK0901 to the University of Stuttgart.
E.L.-R. is supported by the NASA Astrophysics Decadal Survey Precursor Science (ADSPS) Program (NNH22ZDA001N-ADSPS) with ID 22-ADSPS22-0009 and agreement number 80NSSC23K1585.
\end{acknowledgments}

%

\vspace{5mm}


\software{aplpy \citep{aplpy},  
          astropy \citep{astropy:2022,astropy:2018,astropy:2013}
          }


\appendix 


\section{Timescales for internal alignment of dust grains} \label{sec:AligMechInternal}

For completeness, in this section, we explore the relaxation process that leads to the internal alignment of the paramagnetic material assuming silicate dust grains, i.e., Barnett relaxation (\citealt{Purcell1979}). The alignment of the angular momentum, $\textbf{J}$, with the short axis of the grain is referred to as internal alignment, and the Barnett effect is the reverse of the Einstein-de Haas effect (\citealt{1915KNAB...18..696E}). It occurs when grain rotational energy dissipates into heat due to rotating magnetization within a paramagnetic grain rotating around a non-principal axis. The dissipation process will ultimately cause the alignment of the angular velocity and momentum with the short-grain axis, which represents the state of minimum energy. The relaxation time for the Barnett effect in a paramagnetic material (referred to as Barnett relaxation) can be expressed as

\begin{equation}\label{eq:tau_BR}
\tau_{\rm BR,PM} = \frac{\gamma_{e}^{2}I_{\|}^{3}}{VK_{\rm PM}(\omega)h^{2}(h-1)J^{2}} \simeq 0.5\rho_{3}^{2}a^{7}_{-5}f(s_{0.5})\left(\frac{J_{\rm d}}{J}\right)^{2}\left[1+\left(\frac{\omega \tau_{\rm el}}{2}\right)^{2}\right]^{2}~{\rm yr,}
\end{equation}
\noindent
\citep{TH2022}, where $\gamma_{e}=-\frac{g_{e}\mu_{B}}{\hbar}$ is the gyromagnetic ratio of an electron ($g_{e}\simeq 2$ and $\mu_{B}\simeq 9.26\times 10^{-21}\,\rm erg\, G^{-1}$) and $\hbar=\frac{h_{\rm{P}}}{2\pi}$ with $h_{\rm{P}}$ as the Planck's constant, $I_{\|} = \frac{8\pi}{15}\rho s a^{5}$ is the greatest moment of inertia of the dust grain, $\rho$ is the dust grain density, $s = \frac{c}{a}$ is the axial ratio of oblate grains, $c$ is the length of the semi-minor axis and $a$ is the lengths of the semi-major axes (i.e., grain size), $V=\frac{4\pi}{3 a^{3}}$ is the grain volume, $K_{\rm{PM}}(\omega)=\frac{\chi_{2(\omega)}}{\omega}$ for paramagnetic grains \citep{TH2022} with $\chi_{2}(\omega)$ as the magnetic susceptibility and $\omega$ as the grain rotation frequency along the short axis, $h=\frac{I_{\|}}{I_{\perp}}=\frac{2}{(1+s^{2})}$ with the moment of inertia $I_{\perp}=\frac{4\pi}{15}\rho s a (1+s^{2})$, and $J$ is the angular momentum of the grain.

We also show Eq. \ref{eq:tau_BR} in practical units, where $\rho_{3} = \frac{\rho}{3~\rm{gr~cm}^{3}}$, $a_{-5} = \frac{a}{10^{-5}~\rm{cm}}$, $f(s_{0.5})$ is the fraction of Fe atoms in a silicate grains  \citep[e.g., $f=1/7$ for  MgFeSiO$_{4}$ grains;][]{Hoang2014} that can be defined as $f(s_{0.5})=s_{0.5}\left(\frac{1+s_{0.5}^{2}}{2}\right)^{2}$ with $s=\frac{s}{0.5}$ as the ratio of the minor to the major axis, $J_{\rm d}=\sqrt{\frac{I_{\|}k_{\rm B}T_{\rm d}}{(h-1)}}$ is the dust thermal angular momentum with $k_{\rm{B}}$ as the Boltzmann's constant and $T_{\rm{d}}$ as the dust temperature, $\tau_{\rm el}=1.43\times 10^{-10}$ yr is the spin-spin relaxation time of the electron spin. 


If the Barnett relaxation is slower than the randomization of grains caused by gas collisions (Section \ref{subsec:taugas}), the internal alignment of gas particles can be significantly disrupted. Consequently, grains exhibit effective internal alignment when the time required for Barnett relaxation is shorter than the time for gas rotational damping. As a result, the largest possible size for a paramagnetic grain, with its internal alignment along the short axis, can be determined by finding a balance between these two timescales. This relationship is derived in \citealt{2022AJ....164..248H}, such as the maximum grain size for internal alignment is

\begin{eqnarray}
a_{\rm max,int} &\simeq & 0.39\times h^{1/3}St^{1/3} \left(\frac{p_{5.5}^{2}}{\rho_{3}n_{3}T^{1/2}_{\rm{gas,10}}}\right)^{1/6} 
\times \left(\frac{1}{1+(\omega \tau_{\rm tel}/2)^{2}}\right)^{1/3}\left(\frac{(h-1)T_{\rm{gas}}}{T_{\rm{dust}}}\right)^{1/6} \\ \nonumber
&\simeq & 9.12\times10^{4} \omega \rho_{3}^{-2/3} s_{0.5}^{-1/2} p_{5.5}^{1/3}T_{\rm{gas},200}^{7/12}\rho_{3}^{-2/3} 
 a_{-5}^{-5/2}n_{3}^{-1/6}~T_{\rm{dust},15}^{-1/2}h^{1/3}(h-1)^{1/6}
 \left(\frac{1}{1+(\omega \tau_{\rm tel}/2)^{2}}\right)^{1/3}~{\rm \mu m},
\end{eqnarray}
\noindent 
where $p_{5.5} = \frac{p}{5.5}$ is the normalized dipole moment \citep{Draine1996,HL2016a}, $T_{\rm{gas,10}} = \frac{T_{\rm{gas}}}{10~\rm{K}}$ and $T_{\rm{gas,200}} = \frac{T_{\rm{gas}}}{200~\rm{K}}$ are the normalized gas temperatures, $n_{3}=\frac{n_{\rm{H}}}{3~\rm{cm}^{-3}}$ is the normalized gas volume density, and $T_{\rm{dust,15}} = \frac{T_{\rm{dust}}}{15~\rm{K}}$ is the normalized dust temperature.

It can be observed that the maximum internal grain size increases with grain rotation as $St^{1/3}$ and decreases with gas density as approximately $n^{-1/6}_{\rm H}$. For parameters in the inter-arms region, where $T_{\rm gas}=200\,$K, $T_{\rm d}=20\,$K and $n_{\rm H}=1\,\rm cm^{-3}$, the value of $a_{\rm max,int}\sim1.15\,$mm. On the other hand, for the parameters in the arms region, where $T_{\rm dust}=T_{\rm gas}=20\,$K and $n_{\rm H}=10\,\rm cm^{-3}$, the value of $a_{\rm max,int}\sim 0.6\,$mm. Therefore, the standard (sub)micron-sized grain is aligned in such a way that its angular momentum is parallel to the grain's shorter axis, which is commonly referred to as perfect internal alignment.


\bibliography{references}{}
\bibliographystyle{aasjournal}

\end{document}